\documentclass{article}
\usepackage{spconf,amsmath,graphicx}
\usepackage{times}
\usepackage{subfigure}
\usepackage{latexsym,amstext,amsfonts,amssymb,epsfig}
\usepackage{algorithm, algorithmic, multirow, xcolor,booktabs}
\usepackage{latexsym}
\usepackage{booktabs}
\usepackage{subfigure}
\usepackage{color}
\usepackage{algorithm}
\usepackage{algorithmic}

\title{Study of Sparsity-Aware Reduced-Dimension Beam-Doppler Space-Time Adaptive Processing}
\name{Zhaocheng Yang$^{1}$, and Rodrigo C. de Lamare$^2$}
\address{ $^1$College of Information Engineering, Shenzhen University, Shenzhen, Guangdong, 518060, China\\
     $^2$Communications Research Group, University of York, York YO10 5DD, United Kingdom\\
     yangzhaocheng@szu.edu.cn and rcdl500@ohm.york.ac.uk}
\begin{document}
%\ninept
%
\maketitle
\begin{abstract}
Existing reduced-dimension beam-Doppler space-time adaptive
processing (RD-BD-STAP) algorithms are confined to the beam-Doppler
cells used for adaptation, which often leads to some performance
degradation. In this work, a novel sparsity-aware RD-BD-STAP
algorithm, denoted Sparse Constraint on Beam-Doppler Selection
Reduced-Dimension Space-Time Adaptive Processing (SCBDS-RD-STAP), is
proposed can adaptively selects the best beam-Doppler cells for
adaptation. The proposed SCBDS-RD-STAP approach formulates the
filter design as a sparse representation problem and enforcing most
of the elements in the weight vector to be zero (or sufficiently
small in amplitude). Simulation results illustrate that the proposed
SCBDS-RD-STAP algorithm outperforms the traditional RD-BD-STAP
approaches with fixed beam-Doppler localized processing.

\end{abstract}
\section{Introduction}

Space-time adaptive processing (STAP) is a leading technology
candidate for improving detection performance of phased-array
airborne radar \cite{Guerci2003} and other related approaches.
However, STAP techniques often suffer from the lack of snapshots for
training the receive filter, especially in nonhomogeneous
environments, which is a crucial concern in the development of STAP
algorithms \cite{Guerci2003,Melvin2006,RuiSTAP2010}.

In the past decades, many related works have been investigated to
improve the clutter mitigation performance in scenarios with a
number of snapshots (see \cite{Guerci2003, Melvin2006, RuiSTAP2010,
Klemm1987, Wang1994, Adve2000.1, wyl2003, Scott1995,
WZhang2014,WangX2016, WangX2015, ZcYangTSP2011, ZcYangEL2017,
ZYangAES2017} and the references therein). For instance, the
auxiliary channel receiver (ACR) \cite{Klemm1987}, the joint domain
localized approach (JDL) \cite{Wang1994, Adve2000.1}, the space-time
multiple-beam (STMB) \cite{wyl2003} are three kinds of effective
reduced-dimension (RD) algorithms in the beam-Doppler domain.
However, the filter design in \cite{Klemm1987, Wang1994, Adve2000.1,
wyl2003} relies on fixed beam-Doppler cells and cannot provide
optimal selection, suffering significant performance degradation in
the presence of sensor array errors. To overcome this issue, the
studies in \cite{Scott1995} and \cite{WZhang2014} proposed
sequential methods that reduce the required partially adaptive
dimension in the transformed domain.

Motivated by the rank deficiency in clutter suppression,
sparsity-aware beamformers have been proposed to improve the
convergence by exploiting the sparsity of the received data and
filter weights \cite{ZcYangTSP2011,ZcYangEL2017}. The studies in
\cite{WangX2015} and \cite{WangX2016} developed a Min-Max STAP
strategy based on the selection of an optimum subset of
antenna-pulse pairs that maximizes the separation between the target
and the clutter trajectory. Both the sparsity-aware beamformers and
the Min-Max STAP strategy are in the antenna-pulse domain. The
former is a data-dependent strategy and the latter is a
data-independent strategy which requires prior knowledge of the
clutter ridge. By drawing inspiration from compressive sensing,
recently reported sparsity-based STAP algorithms have formulated the
STAP problem as a sparse representation that exploits the sparsity
of the entire observing scene in the whole angle-Doppler plane
\cite{ZYangAES2017}. However, this kind of approach suffers from
high computational complexity due to the large dimension of the
discretized angle-Doppler plane. Previous works imply that the
degrees of freedom (DoFs) used for STAP filters required to mitigate
the clutter are much smaller than the full dimension, and different
selection strategies have resulted in various levels of performance.

In this work, we introduce the idea of sparse selection in the
beam-Doppler domain and formulate the STAP filter design as a sparse
representation problem. Unlike the sparsity-based STAP
\cite{ZYangAES2017}, the proposed Sparse Constraint on Beam-Doppler
Selection Reduced-Dimension STAP (SCBDS-RD-STAP) algorithm does not
discretize the angle-Doppler plane into a large number of grids, but
only transforms the received data into a same size beam-Doppler
domain. Differently from the sparsity-aware beamformers
\cite{ZcYangTSP2011,ZcYangEL2017} or the Min-Max STAP strategy
\cite{WangX2015, WangX2016}, the proposed SCBDS-RD-STAP algorithm
designs the filter in the beam-Doppler domain and automatically
selects the best beam-Doppler cells used for adaptation by solving a
sparse representation problem. In addition, an analysis of the
complexity is performed for the proposed algorithm. Simulation
results show the effectiveness of the proposed algorithm.

This paper is structured as follows: Section 2 describes the signal
model of a pulse Doppler side-looking airborne system and states the
problem. Section 3 details the proposed SCBDS-RD-STAP algorithm
along with approximate solutions and their computational complexity.
Section 4 presents and discusses simulation results while Section 5
provides the concluding remarks.

\section{Signal Model and Problem Formulation}

In this section we describe the signal model of a pulse Doppler
side-looking airborne radar system and state the problem of
designing a beam-Doppler STAP.

\subsection{Signal Model}

Considering a pulse Doppler side-looking airborne radar with a
uniform linear array (ULA) consisting of $M$ elements. The radar
transmits a coherent burst of $N$ pulses at a constant pulse
repetition frequency (PRF) $f_r$. Generally, for a range bin with
the space-time snapshot ${\bf x}$, target detection can be
formulated as a binary hypothesis problem and expressed as
\begin{eqnarray}\label{eq1}
    H_0: {\bf x}& = &{\bf x}_u \nonumber\\
    H_1: {\bf x}& = &\alpha_t{\bf s}+{\bf x}_u,
\end{eqnarray}
where $H_0$ and $H_1$ denote the disturbance only and the target
plus disturbance hypotheses, respectively, $\alpha_t$ is a complex
gain, ${\bf s}$ is the $NM \times1$ target space-time steering
vector and ${\bf x}_u$ denotes the clutter-plus-noise vector which
encompasses the clutter and the thermal noise \cite{Guerci2003}.

The STAP filter based on a minimum variance distortionless response
(MVDR) approach by minimizing the clutter-plus-noise output power
while constraining a unitary gain in the direction of the desired
target signal is expressed as \cite{RuiSTAP2010}
\begin{equation}\label{eq4}
    {\bf w}_{\rm opt}=\frac{{\bf R}^{-1} {\bf s}}{{\bf s}^H{\bf R}^{-1} {\bf s}}.
\end{equation}
where ${\bf R} = E\left[{\bf x}_u {\bf x}^H_u \right]$ denotes the
clutter-plus-noise covariance matrix. Approaches to compute the
beamforming weights include
\cite{xutsa,delamaretsp,kwak,xu&liu,spa,delamareccm,wcccm,delamareelb,jidf,delamarecl,delamaresp,delamaretvt,jioel,rrdoa,delamarespl07,delamare_ccmmswf,jidf_echo,delamaretvt10,delamaretvt2011ST,delamare10,fa10,lei09,ccmavf,lei10,jio_ccm,ccmavf,stap_jio,zhaocheng,zhaocheng2,arh_eusipco,arh_taes,dfjio,rdrab,locsme,okspme,dcg_conf,dcg,dce,drr_conf,drr_j,dta_conf1,dta_conf2,dta_ls,damdc,song,wljio,barc,jiomber,saalt}

\subsection{The Beam-Doppler STAP Approaches}

The beam-Doppler STAP approaches firstly transform the data ${\bf
x}$ in the antenna-pulse domain to the beam-Doppler domain, denoted
as $\tilde{\bf x}$, where `` $\tilde{}$ " above ${\bf x}$ signifies
the beam-Doppler domain. This procedure can be represented by
\begin{equation}\label{eq5}
    \tilde{\bf x} = {\bf T}^H_{\rm LP} {\bf x},
\end{equation}
where ${\bf T}_{\rm LP}$ denotes the transformation matrix. The
common idea under the beam-Doppler STAP approaches is to choose a
localized processing (LP) region, or equivalently, the matrix ${\bf
T}_{\rm LP}$, corresponding to a set of beam-Doppler responses, for
adaptive processing. The optimal beam-Doppler STAP filter can be
represented by
\begin{equation}\label{eq10}
    \tilde{\bf w}_{\rm opt} = \frac{\tilde{\bf R}^{-1} \tilde{\bf s}}{\tilde{\bf s}^H\tilde{\bf R}^{-1} \tilde{\bf s}},
\end{equation}
where $\tilde{\bf R}={\bf T}^H_{\rm LP}{\bf R}{\bf T}_{\rm LP}$ and $\tilde{\bf s}={\bf T}^H_{\rm LP}{\bf s}$.

\begin{figure}[ht]
  \centering
  \subfigure[ACR]{\label{ACR}
  \includegraphics[width=2.7cm]{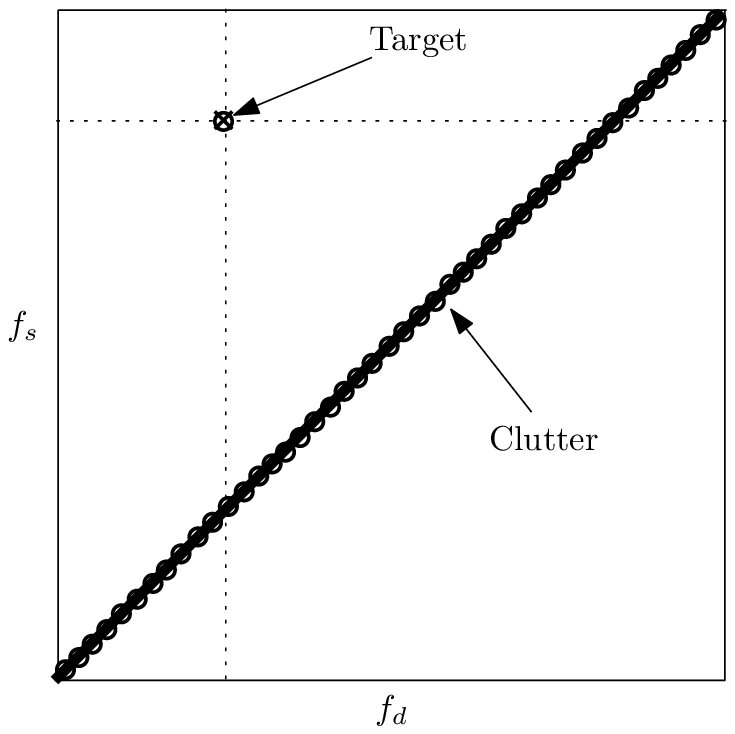}}
  %\hspace{0.8in}
  \subfigure[JDL]{\label{JDL}
  \includegraphics[width=2.7cm]{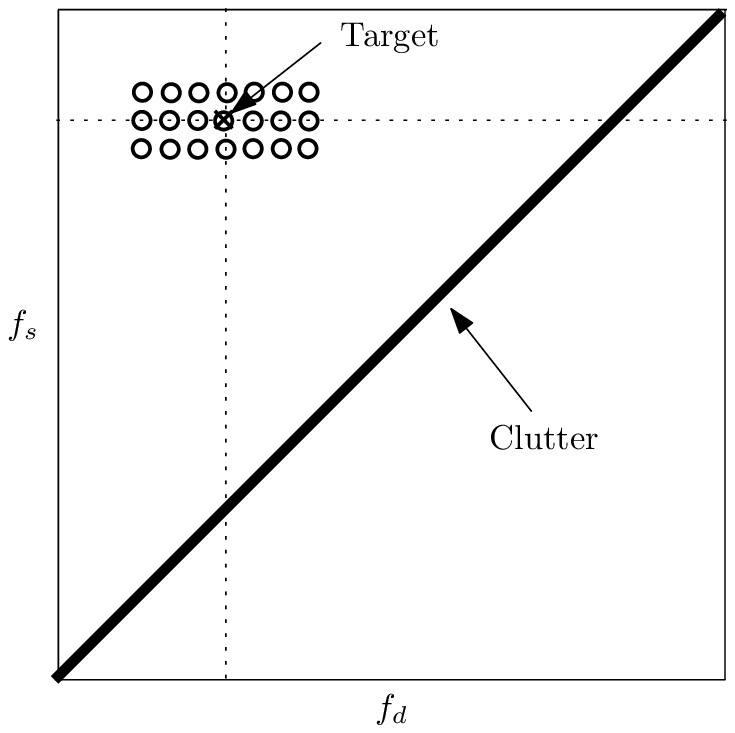}}
  %\hspace{0.3in}
  \subfigure[STMB]{\label{STMB}
  \includegraphics[width=2.7cm]{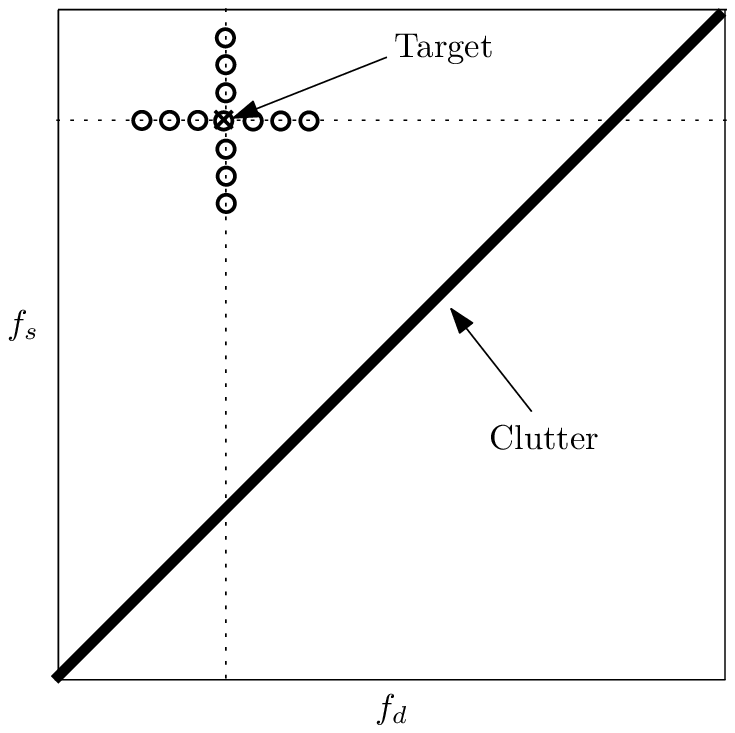}}
  \caption{The LP region selections in different beam-Doppler LP approaches. $\circ$ denotes the selected beam-Doppler cell, and $\times$ denotes the target beam-Doppler cell.}\label{LPR_Select}
\end{figure}

Observing (\ref{eq10}), the key challenge is how to efficiently
select the LP region. The ACR method \cite{Klemm1987} suggests to
select the LP region placed along the clutter ridge, as shown in
Fig.\ref{ACR}. The JDL method \cite{Wang1994,Adve2000.1} chooses the
beam-Doppler cells around the target cell, which turns out to be a
rectangular shape, as shown in Fig.\ref{JDL}. Unlike ACR and JDL,
STMB \cite{wyl2003} chooses the beam-Doppler cells with a ``cross"
shape centered at the target cell, as shown in Fig.\ref{STMB}. All
these approaches can reduce the STAP filter dimension, resulting in
improved convergence and steady-state performance in a small
training data set. However, the ACR requires the knowledge of the
clutter ridge, and there is no rule to determine the optimum size of
the chosen beam-Doppler LP region for JDL and STMB. The optimum
choice of the beam-Doppler region should be related to the scenario
or the data rather than just fixed.

\section{Proposed SCBDS-RD-STAP Algorithm}

In this section, we detail the proposed SCBDS-RD-STAP algorithm, how
to design the receive filter and discuss the computational
complexity.

\subsection{Proposed SCBDS-RD-STAP Scheme}
\label{scbds-stap}

The core idea of the proposed SCBDS-RD-STAP scheme is based on a
transformation matrix and a filter with sparse constraints. The
received space-time data vector ${\bf x}$ is first mapped by an $NM
\times NM$ transformation matrix ${\bf T}$ into an $NM \times 1$
beam-Doppler domain data vector. Here, ${\bf T}$ can be constructed
as
\begin{eqnarray}\label{eqsc0}
    {\bf T} =
    \left[\begin{array}{cc} {\bf s}  & {\bf T}_{\rm aux} \end{array}\right],
\end{eqnarray}
where ${\bf T}_{\rm aux}$ is an $NM \times (NM-1)$ matrix, given by
\begin{eqnarray}\label{eqsc0.1}
    {\bf T}_{\rm aux} =
     \left[\begin{array}{c} \left({\bf v}_d(f_{d,t}) \otimes {\bf v}_s(f_{s,t} + \frac{1}{M})\right)^T \\
    \vdots \\
    \left({\bf v}_d(f_{d,t}) \otimes {\bf v}_s(f_{s,t} + \frac{M-1}{M})\right)^T  \\
    \left({\bf v}_d(f_{d,t}+\frac{1}{N}) \otimes {\bf v}_s(f_{s,t})\right)^T  \\
    \vdots \\
    \left({\bf v}_d(f_{d,t} + \frac{1}{N}) \otimes {\bf v}_s(f_{s,t} + \frac{M-1}{M})\right)^T  \\
    \vdots \\
    \left({\bf v}_d(f_{d,t} + \frac{N-1}{N}) \otimes {\bf v}_s(f_{s,t} + \frac{M-1}{M})\right)^T
    \end{array}\right]^T.
\end{eqnarray}

Denoting $d={\bf s}^H{\bf x}$ and $\tilde{\bf x}={\bf T}_{\rm aux}^H
{\bf x}$, we note that $d$ is the component at the target
beam-Doppler cell (also called main channel), and elements of
$\tilde{\bf x}$ are the components from otherwise beam-Doppler cells
(also called auxiliary channels). Following the concept of GSC, we
can expect to reduce the clutter in $d$ by employing a filter on the
auxiliary channel data $\tilde{\bf x}$. Furthermore, based on the
first three observations analyzed above, we do not need to use all
auxiliary channel data but only a few of them. In order to realize
this idea, we perform a sparse constraint on the STAP filter weight
vector $\tilde{\bf w}$. Precisely, we design the filter $\tilde{\bf
w}$ by solving the following optimization problem
\begin{eqnarray}\label{eqsc3}
\begin{split}
    \min_{\tilde{\bf w}} E\left[\left|d - \tilde{\bf w}^H\tilde{\bf x}\right|^2\right] + \kappa \left\|\tilde{\bf w}\right\|_0,
\end{split}
\end{eqnarray}
where $\kappa$ is the regularization parameter that controls the
balance between the sparsity and total squared error. Theoretically,
the optimum choice can be determined by an algorithm that is
properly designed for the task. To show an intuitive observation of
the above idea, we will provide examples by simulations later on. I
am not sure about the above but it would be useful to include a
table with the pseudo-code of the SCBDS-RD-STAP algorithm here.

\subsection{Approximate Solutions}

Since the sparse regularization function is $l_0$-norm, it leads to
an NP-hard problem. In the following, we use the relaxation penalty
$l_p$-norm (where $0<p\leq 1$) instead of the $l_0$-norm and rewrite
(\ref{eqsc3}) as
\begin{eqnarray}\label{eqsc4}
    \min_{\tilde{\bf w}} E\left[\left|d - \tilde{\bf w}^H\tilde{\bf x}\right|^2\right] + \kappa \left\|\tilde{\bf w}\right\|_p.
\end{eqnarray}

In practice, since the expectation in (\ref{eqsc4}) cannot be
obtained, we now modify (\ref{eqsc4}) based on a least-squares type
cost function. Let ${\bf X}=\left[{\bf x}_1, {\bf x}_2, \cdots, {\bf
x}_L\right]$ denote the space-time data matrix formed by $L$
training snapshots, and let ${\bf d}={\bf X}^T{\bf s}^{\ast}$, and
$\tilde{\bf X}={\bf T}_{\rm aux}^H {\bf X}$, then the least-squares
type cost function is described by
\begin{eqnarray}\label{eqsc6}
    \min_{\tilde{\bf w}} \left\|{\bf d}^\ast - \tilde{\bf X}^H\tilde{\bf w}\right\|^2_2 + \kappa \left\|\tilde{\bf w}\right\|_p.
\end{eqnarray}
%\begin{eqnarray}\label{eqsc5}
%    \min_{\tilde{\bf w}} \left\|{\bf d} - \tilde{\bf X}^T\tilde{\bf w}^\ast\right\|^2_2 + \kappa \left\|\tilde{\bf w}\right\|_p.
%\end{eqnarray}
Note that (\ref{eqsc6}) is a standard sparse representation problem
and can be solved by the regularized focal underdetermined system
solution (R-FOCUSS) algorithm.

It should be noted that the sensing matrix or the dictionary
$\tilde{\bf X}^H$ of the optimization problem (\ref{eqsc6}) of the
proposed SCBDS-RD-STAP scheme is formed by the received data (i.e.,
snapshots from the beam-Doppler domain), and is different from that
of the sparsity-based STAP approaches \cite{ZYangAES2017}, which is
composed of known space-time steering vectors from the discretized
angle-Doppler plane. Furthermore, unlike the ACR, JDL, and STMB,
which are performed with fixed beam-Doppler LP region, the proposed
SCBDS-RD-STAP scheme provides an iterative approach to automatically
select the beam-Doppler LP region aided by a sparse constraint.
Additionally, the auxiliary channel data are formulated by a
standard 2-D discrete Fourier transform with explicit physical
meaning in the proposed SCBDS-RD-STAP scheme, whereas the auxiliary
channel data are formulated by a signal blocking matrix in the
sparsity-aware beamformer \cite{ZcYangTSP2011}.

\subsection{Computational Complexity }

We detail the computational complexity of the proposed SCBDS-RD-STAP
algorithm, sparsity-aware beamformer \cite{ZcYangTSP2011}, and JDL
\cite{Adve2000.1}/STMB \cite{wyl2003}, as shown in Table
\ref{complexity}.
%$O\big%(\sum^{K_{\rm foc}}_{q=1}D_{{\rm foc},q}L^2\big)$, $O\big(5L(NM)^2\big)$,
%and $O\big(LD^2 + D^3\big)$, respectively.
Here, for the proposed algorithm, $K_{\rm foc}$ is the total
iteration number and $D_{{\rm foc},q}$ is the number of elements
above the preset threshold at the $q$th iteration, which is decided
by the sparsity; for the JDL/STMB, $D$ is the number of selected
beam-Doppler elements. From the table, we see that the computational
complexity of the proposed algorithm is comparable or even
lower\footnote{In our experiments, the average time used for the
proposed algorithm with a fixed $p$ and the sparsity-aware
beamformer are $0.033$ second and $0.1$ second, respectively.} than
that of the sparsity-aware beamformer, and higher than those of the
JDL and STMB. This is because the number of snapshots $L$ used in
the proposed SCBDS-RD-STAP algorithm is much smaller than $NM$
(which can be seen in the simulations), the value of $D_{{\rm
foc},q}$ after several iterations will also be much smaller than
$NM$, and the pseudo-inversion can be calculated by the conjugate
gradient approach, which has low complexity \cite{ZHe2009}.

\begin{table}[!ht]
  \centering
  \caption{Computational Complexity}\label{complexity}

  \begin{tabular}{|l|c|c|}
  \hline
  Algorithm & Complexity \\
  %\hline
%   GSC-SMI & $O\left((NM-1)^3+L(NM)^2\right)$\\
  \hline
   Sparsity-aware beamformer & $O\left(5L(NM)^2\right)$\\
  \hline
   JDL/STMB &$O\left(LD^2 + D^3\right)$\\
  \hline
   Proposed SCBDS-RD-STAP &$O\left(\sum^{K_{\rm foc}}_{q=1}D_{{\rm foc},q}L^2\right)$\\
  \hline
   \end{tabular}
\end{table}

\section{Simulations}

In this section, we assess the performance of the proposed
SCBDS-RD-STAP algorithm and compare it with other existing
algorithms, namely, the JDL ($3 \times 3$) \cite{Adve2000.1}, STMB
($8 + 4 + 1$) \cite{wyl2003}, and sparsity-aware beamformer
\cite{ZcYangTSP2011} in terms of the output
signal-to-clutter-plus-noise-ratio (SCNR) loss \cite{Guerci2003},
which is defined as
\begin{eqnarray}\label{det2}
    {\rm SCNR}_{\rm loss} = \frac{\sigma^2\left|\hat{\bf w}^H{\bf s}\right|^2}{NM\hat{\bf w}^H{\bf R}\hat{\bf w}},
\end{eqnarray}
where $\hat{\bf w} = {\bf s} - {\bf T}_{\rm aux}\tilde{\bf w}$ is
the corresponding filter weight vector in the original domain. We
consider a side-looking ULA (half-wavelength inter-element spacing)
airborne radar with the following parameters: uniform transmit
pattern, $M=12$, $N=12$, carrier frequency $1.2$GHz, $f_r=2$kHz,
platform velocity $125$m/s, platform altitude $8$km,
clutter-to-noise ratio (CNR) $45$dB. For the following examples: in
the sparsity-aware beamformer, we set parameters as those in
\cite{ZcYangTSP2011}; in the proposed SCBDS-RD-STAP algorithm, we
set the regularization parameter to $3$, the maximum iteration
number is $500$, and the stopping criterion is decided by the preset
limit relative change of the solution between two adjacent
iterations $10^{-4}$.

\begin{figure}[!htb]
\centering
\includegraphics[width=8cm,height=6cm]{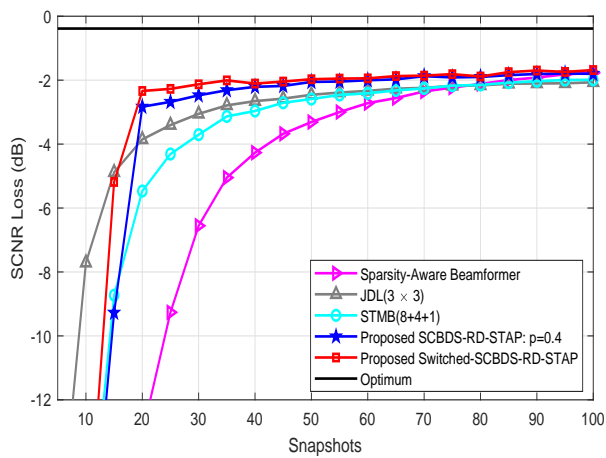}
\caption{The SCNR loss against the number of snapshots for training.} \label{sinr_snapshots}
\end{figure}

In the first example, we examine the convergence performance
(signal-to-clutter-plus-noise ratio (SCNR) loss against the number
of snapshots) of the proposed SCBDS-RD-STAP algorithm, as shown in
Fig.\ref{sinr_snapshots}. The true target is supposed to be
boresight aligned with normalized Doppler frequency $-0.1667$. The
curves show that the proposed Switched-SCBDS-RD-STAP algorithm
converges to a higher SCNR loss with much fewer training snapshots
compared to all the considered algorithms.

\begin{figure}[!htb]
\centering
\includegraphics[width=8cm]{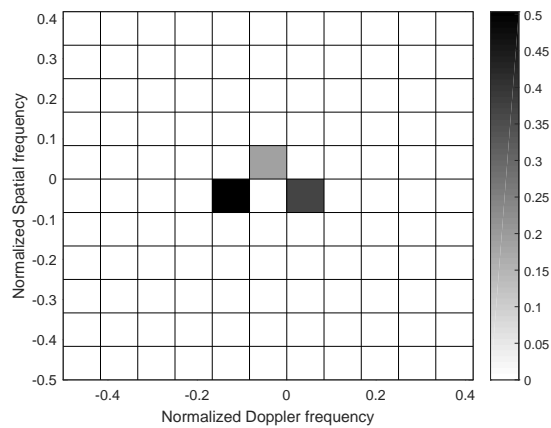}
\caption{2-D view of the weight vector of the Switched-SCBDS-RD-STAP}
\label{selected_bd_cells}
\end{figure}

Fig.\ref{selected_bd_cells} illustrates the 2-D view of the weight
vector, specifically, each element in the weight vector is
represented by one grid point, and its amplitude is depicted by the
grayscale of the grid. Note that, each element in the weight vector
is associated to one auxiliary channel in the GSC, and a zero
amplitude implies the associated auxiliary channel is not involved
in the adaption. Apparently, most of the elements in the weight
vector have zero amplitudes, which implies that the
Switched-SCBDS-RD-STAP selects very few beam-Doppler cells for
adaptation.

\begin{figure}[ht]
  \centering
  \includegraphics[width=8cm,height=6cm]{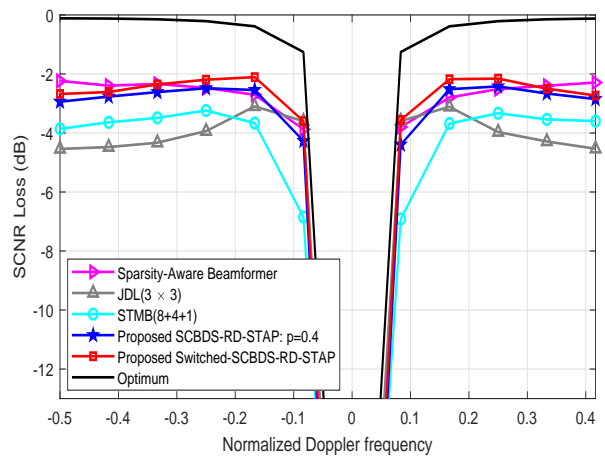}
  \caption{The SCNR loss versus different target Doppler frequencies.}\label{SINR_Dop}
\end{figure}

In the third example, we assess the performance of the proposed
SCBDS-RD-STAP algorithm under different target Doppler frequencies,
as depicted in Fig.\ref{SINR_Dop}. Here, we set the number of
snapshots for training used in the JDL ($30$), STMB ($30$), proposed
SCBDS-RD-STAP ($30$), and sparsity-aware beamformer ($60$). The
curves illustrates that the proposed SCBDS-RD-STAP algorithm
provides much better performance than other algorithms for small
target Doppler frequencies. That is to say, the proposed
SCBDS-RD-STAP algorithm is suitable for slow moving target
detection. Although the performance of the SCBDS-RD-STAP algorithm
is slightly lower than that of the sparsity-aware beamformer for
large target Doppler frequencies, the number of snapshots used in
the SCBDS-RD-STAP algorithm is much less.

\section{Conclusions}

This paper has proposed a novel STAP algorithm based on the
beam-Doppler selection for clutter mitigation for airborne radar
with small sample support. The SCBDS-RD-STAP algorithm transforms
the received data into beam-Doppler domain, employs a sparse
constraint on the filter weight for sparse beam-Doppler selection
and formulates this selection as a sparse representation problem,
where the sensing matrix is formed by the data matrix. Simulations
have demonstrated the effectiveness of the proposed SCBDS-RD-STAP
algorithm and shown its improvement in target detection over the
existing algorithms, such as the JDL, STMB, and sparsity-aware
beamformer both in absence and presence of array errors.

\end{document}